\pdfoutput=1
\documentclass[11pt]{article}
\usepackage[T1]{fontenc}
\usepackage[utf8]{inputenc}
\usepackage{acl}
\usepackage{times}
\usepackage{latexsym}
\usepackage{booktabs}
\usepackage{amsmath}
\usepackage{graphicx}

\title{PageRecall: Measuring Page Selection in\\
Literature-Grounded Question Answering}

\author{Aaditya Chauhan \\
  Independent Researcher \\
  Team: Everest \\
  \texttt{aadityc28@gmail.com}}

\begin{document}
\maketitle

\begin{abstract}
We describe our system for LitTraceQA (GroundLM @ EMNLP 2026): given a research
question, retrieve the relevant papers from a pool of 27,487, cite the page and the table or figure where the answer lives, and answer in a
requested format. Our main finding
is that evidence grounding is limited by \emph{retrieval}, not by reading. The page selector put the annotator's page, which we call the gold page, in
front of the model that locates evidence only about half the time (52.6\% gold-page
recall), while that model, given the page, cited the right one in 45 of the 48
locators it emitted (94\%). When the page was missing it rarely said so: of 45 such cases it returned nothing 14 times, a wrong page 24 times, and a
correct page 7 times, so the pipeline failed quietly almost twice as often as it
failed visibly. Since the failure was that the right page was never shown, the fix is to stop
choosing: each retrieved paper fits in the model's context, so we show it whole.
Page ranking survives only as a fallback inside papers too long to fit, which no
test-split paper was, and gold-page recall reaches 100\% on the papers we can
parse. Separately, questions that identify their target
by position rather than content, such as ``the first author of the 24th
reference'', are served by parsing rather than retrieval: we resolve the
bibliography into an addressable list, which also supplies identifiers the
evidence metric scores. The final system scores 0.762 paper $F_1$, 0.441 evidence $F_1$ and 0.920
multiple-choice accuracy on the held-out test split. Because the pipeline depends
on a closed model without seed control, we release a harness that verifies the paper's central claims against committed
artifacts.
\end{abstract}

\section{Introduction}

LitTraceQA \cite{liu2026littraceqabenchmarkmultistagegrounding} asks a system to do three things at once: retrieve the relevant
paper(s) from a released metadata pool, identify \emph{coarse evidence
locations} (page, plus table or figure number), and produce an answer in a
requested format. Evidence grounding is scored separately from answer accuracy,
which lets us separate two failures that are normally confounded. When a
grounded QA system cites the wrong place, is the fault with the model that reads
a paper and picks the citation, which we call the \emph{reader}, or with the step
that chose which pages the reader was allowed to see?

On this task the answer is the selector.

Our contributions are:
\begin{itemize}
\item A controlled measurement isolating page selection from paper retrieval
  (\S\ref{sec:pageselect}): over 97 units the selector showed the reader the
  gold page only about half the time (52.6\% gold-page recall), while the
  reader, given the gold page, cited the right page in 45 of the 48 locators it
  emitted (94\%). Without the gold page it rarely abstained: in 31 of 45 units
  it emitted a locator anyway, and only 7 of those 31 were correct.
\item A cautionary result on tuning a system to a development split: we fixed
  how many papers to predict per question from the development distribution, and
  it transferred badly (\S\ref{sec:setsize}).
\item A positional bias in multiple-choice answering that aggregate accuracy
  hides but the predicted-label distribution exposes, and a permutation-voting
  correction for it (\S\ref{sec:mcbias}).
\item A pre-emit entailment gate (\S\ref{sec:gate}) that audits answers against
  their own cited source text, and the negative result that groundedness does
  not predict correctness on this data.
\end{itemize}

\section{Related Work}

The closest prior work scores evidence selection over scientific papers
directly. QASPER \cite{dasigi2021qasper} scores evidence-paragraph selection alongside
answering, and isolates the two with oracle experiments that hand models the
gold evidence, the same decomposition our conditional measurement makes at page
level. Context24 \cite{chan2024context24} targets figure- and table-level evidence for
scientific claims, the granularity our caption index serves; a participant system there likewise used captions as a retrieval
representation \cite{bolucu2024csiro}.

For locating evidence inside long documents, one line of work retrieves
\emph{page images} directly using late interaction \cite{khattab2020colbert},
notably ColPali \cite{faysse2025colpali} and M3DocRAG
\cite{cho2024m3docrag}, which sidestep text extraction entirely. These are the
natural stronger baselines for our selector, and we discuss why we did not adopt
them in \S\ref{sec:pageselect}. LongDocURL \cite{deng2025longdocurl} scores evidence locating explicitly and
MMLongBench-Doc \cite{ma2024mmlongbench} annotates evidence pages; both would be
suitable external validation for this component.

Our \emph{structural anchors} (\S\ref{sec:anchors}), deterministic lookups that
resolve a question naming ``Table 3'' or ``the 24th reference'' against the
parsed document, rely on document parsing rather than retrieval. GROBID \cite{lopez2009grobid} and Docling \cite{auer2024docling} produce full
structured representations of a document. Ours is far lighter, a set of regular
expressions over extracted text, and a real parser would likely do better. Finally, feeding a whole document instead of selected
passages is only viable if long-context models use their context uniformly;
\citet{liu2024lost} show position-dependent degradation, which we discuss as the
main threat to our chosen fix. Query-side rewriting such as HyDE
\cite{gao2023hyde} addresses the vocabulary-mismatch subset of our failures and
remains untried.

\section{Task and Data}

The benchmark paper \cite{liu2026littraceqabenchmarkmultistagegrounding} and the
shared-task findings paper \cite{wang-etal-2026-groundlm} describe the data; we
note only what our design turns on. LitTraceQA is one of two GroundLM 2026 shared
tasks; the other, GoldenViewVQA \cite{wang2026doesanswercomefrom}, asks for
view-level visual evidence in multi-view driving scenes and is not addressed here. The pool is 27{,}487 papers (2024--2025) across nine venues, with title,
abstract, venue, year and a PDF URL but no full text. Development is 55 labelled
questions, the scored test split 71, and neither challenge split contains
freeform questions. Evidence is scored on a coarse key of paper, source type,
page and a normalised object id, so a correct object cited on the wrong page
scores nothing. Most consequentially, \textbf{the inference-time input does not
state the task family}: gold records carry a \texttt{task\_family} field and
released inputs do not, so family must be inferred from the question before any
retrieval policy can be chosen. That single omission shapes the whole
pipeline.

\section{System}

\begin{figure}[t]
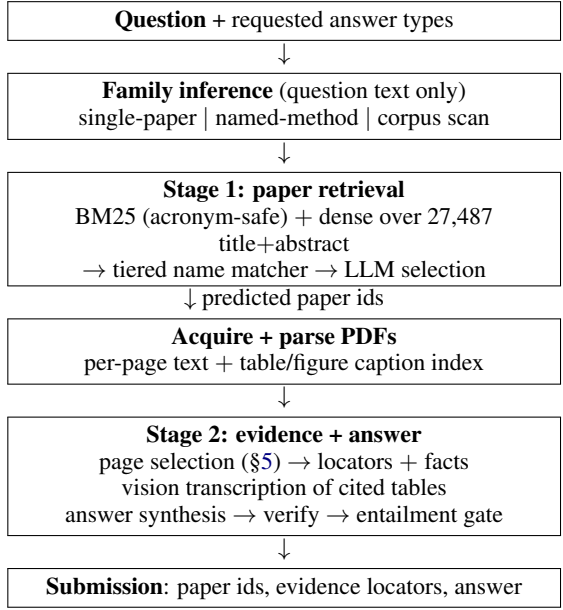

\centering
\small
\setlength{\fboxsep}{3pt}
\begin{tabular}{@{}c@{}}
\fbox{\begin{minipage}{0.92\columnwidth}\centering
\textbf{Question} + requested answer types
\end{minipage}}\\[2pt]
$\downarrow$\\[2pt]
\fbox{\begin{minipage}{0.92\columnwidth}\centering
\textbf{Family inference} (question text only)\\
\footnotesize single-paper $\mid$ named-method $\mid$ corpus scan
\end{minipage}}\\[2pt]
$\downarrow$\\[2pt]
\fbox{\begin{minipage}{0.92\columnwidth}\centering
\textbf{Stage 1: paper retrieval}\\
\footnotesize BM25 (acronym-safe) $+$ dense over 27{,}487 title$+$abstract\\
\footnotesize $\rightarrow$ tiered name matcher $\rightarrow$ LLM selection
\end{minipage}}\\[2pt]
$\downarrow$ \footnotesize predicted paper ids\\[2pt]
\fbox{\begin{minipage}{0.92\columnwidth}\centering
\textbf{Acquire + parse PDFs}\\
\footnotesize per-page text $+$ table/figure caption index
\end{minipage}}\\[2pt]
$\downarrow$\\[2pt]
\fbox{\begin{minipage}{0.92\columnwidth}\centering
\textbf{Stage 2: evidence + answer}\\
\footnotesize page selection (\S\ref{sec:pageselect}) $\rightarrow$ locators $+$ facts\\
\footnotesize vision transcription of cited tables\\
\footnotesize answer synthesis $\rightarrow$ verify $\rightarrow$ entailment gate
\end{minipage}}\\[2pt]
$\downarrow$\\[2pt]
\fbox{\begin{minipage}{0.92\columnwidth}\centering
\textbf{Submission}: paper ids, evidence locators, answer
\end{minipage}}
\end{tabular}
\caption{The pipeline. Family is inferred from the question alone because the
inference-time input does not state it. Stage 1 selects papers; stage 2 locates
evidence inside them and answers.}
\label{fig:pipeline}
\end{figure}

Figure~\ref{fig:pipeline} shows the pipeline end to end; the rest of this
section describes each stage in turn.

\paragraph{Models, tools and compute.}
All language and vision calls go to \textbf{\texttt{claude-sonnet-5}}, accessed
through the Claude command-line interface (version recorded per run; 2.1.224 for
the submitted results) with the model pinned on every call. Decoding uses the
interface defaults and exposes no seed or temperature control, which is why the
response cache rather than the source is what makes a run reproducible
(\S\ref{sec:repro}). The same model performs table and figure transcription, given
a rendered page image. Nothing is fine-tuned and we release no checkpoints.
Dense retrieval uses \texttt{Snowflake/\allowbreak snowflake-\allowbreak arctic-\allowbreak embed-m-v1.5} unmodified
on CPU; sparse retrieval uses \texttt{bm25s} with PyStemmer; PDF parsing uses
\texttt{pymupdf}/\texttt{pymupdf4llm} 1.28. External data is limited to the released paper-metadata pool and paper PDFs
fetched at run time from publisher and proceedings sources; those PDFs are not
redistributed. No synthetic or generated training data is used. The response
cache is released separately from the code because its vision entries contain
verbatim transcriptions of tables from copyrighted papers.
One full run of the pipeline over both splits consumes 965 model calls in seven
categories, measured by replaying it against the cache: evidence location 222,
option-order voting 182, answer synthesis 126, verification 126, vision
transcription 117, entailment auditing 97 and retrieval selection 95; a further
64 question-level calls resolve the corpus-scan questions, for 1{,}029 in all. The released cache
holds 2{,}012 entries; the excess is superseded configurations accumulated
during development. Throughput, measured from response-cache write times over the three largest
recorded runs (298--365 fresh calls each), was one completed call every three to
five seconds in aggregate with four concurrent workers, so a full run over both
splits costs roughly an hour to an hour and a half of wall-clock time; retrieval
indexes are built once on CPU. No GPU is used at any stage.

\paragraph{Stage 1: paper retrieval.}
We build two indexes over title$+$abstract. Sparse retrieval uses BM25
\cite{robertson2009bm25} via \texttt{bm25s} with an acronym-preserving
tokenizer: stemming destroys method names such as \texttt{D-FINE} or
\texttt{sCM}, which are the query terms that matter here. Documents
additionally index \emph{title-initialism aliases} (e.g.\ \texttt{tcm} for
``Truncated Consistency Models''), because a paper's acronym self-name usually
appears only in its full text, never in its metadata. The dense index embeds the
same title and abstract text with arctic-embed-m, unmodified. 

Family inference then routes each question. Single-paper questions go to an LLM
selection over a blended candidate list. Multi-paper questions, mostly
named-method lookups, are resolved by a tiered name matcher (verbatim in
title; verbatim in abstract; letters an in-order subsequence of title-word
initials; letters a multiset subset thereof) followed by an LLM that maps each
named method to the paper introducing it. Corpus-scan questions (``which CVPR
2025 papers cite X'') get a venue/year filter, an LLM abstract screen, then a
full-text check of the shortlist.

\paragraph{PDF acquisition.}
Roughly 44\% of the pool carries an OpenReview URL that blocks scripted fetches, so we try the publisher URL, then the official proceedings mirror, and only then
arXiv. The order matters because a preprint is a different document with
different pagination, and page is part of the scored evidence key: on development, gold table and figure pages agree with our parse in 59 of 82
locators (72\%) on official copies, 33 of 48 for papers fetched from the
proceedings mirrors and 26 of 34 from publisher PDFs, against about 41\% on
preprints in an earlier measurement. On the test split every predicted paper was acquired, and all
42 OpenReview-blocked papers resolved to official copies.

\paragraph{Parsing.}
We parse location-preservingly: per-page text in correct reading order via
\texttt{pymupdf4llm}, plus a table/figure caption index giving each object an id
and a page. The atomic unit is the physical PDF page, which exactly matches the
granularity the evidence metric scores, there is no chunk-to-page mapping to
get wrong.

\paragraph{Stage 2: evidence and answers.}
For each predicted paper an LLM sees the caption index and the paper text, and
returns evidence locators plus extracted facts. Cited tables and figures are
re-read by a vision model from a rendered page image, and every numeric token
the model emits is checked against the page's raw character set by exact string match, a cheap
mechanical check for invented digits. It cannot catch a value that is wrong but
present elsewhere on the page, nor judge semantic equivalence. A synthesis call then produces the answer, a
conservative verification pass may correct it, and the entailment gate
(\S\ref{sec:gate}) audits it. \texttt{validate\_evidence} drops any locator
whose object id does not exist in the parse.

\section{Page selection is the bottleneck}
\label{sec:pageselect}

\paragraph{Setup.}
We want to measure the page selector alone, without confounding it with
mistakes made earlier by paper retrieval. The unit of evaluation is therefore a
\emph{(question, paper)} pair rather than a question: a development question
citing three gold papers contributes three units, and units are kept only where
stage 1 already retrieved that paper and we hold a parse of it. The 55
development questions yield 97 such units. Ground truth is the set of gold
evidence pages for that paper, and the measurement involves no model calls, so it
is deterministic and free to repeat.

\paragraph{Result.}
The original policy (BM25 over pages, top 4) achieves 52.6\% gold-page recall:
the fraction of a unit's gold pages among the four shown, averaged over the 97
units. At least one gold page was shown in 52 of them, and the comparison that
matters is conditional on that:

\begin{center}
\small
\begin{tabular}{lccc}
\toprule
gold page & units & emitted & page correct \\
\midrule
shown     & 52 & 48 & \textbf{45 (94\%)} \\
not shown & 45 & 31 & 7 (23\%) \\
\bottomrule
\end{tabular}
\end{center}

Given the right page, the reader cites it correctly in 45 of the 48 locators it
emits (94\%), so most of the evidence deficit sits upstream of it. Prompt engineering on the reader cannot move a
number gated at 52.6\%.

The second row is its own finding. When the gold page was not among the four pages the original selector showed,
the paper itself being correctly retrieved in every unit, the reader is told it
may return nothing, and mostly does not: it emits a locator in 31 of 45 such
cases and is right in 7 of those 31. The 7 successes are not luck.
All 7 cite a table or a figure (4 figures, 3 tables) and none is a text span,
because the prompt also carries a caption index listing every table and figure
in the paper with its page, whether or not that page's text was selected; the
index is there so the model can cite the object identifiers the evidence metric
scores. A model
can therefore cite ``Figure 3'' from the index having never read the page, and
our locator validation then snaps the citation to the page where that object
actually sits. Page selection thus gates text-span evidence hardest, while
table and figure evidence has a second channel that partly bypasses it. The
practical consequence is that this pipeline failed quietly about twice as often
as it failed visibly, and inspecting only the cases where the model returned
nothing would have found the smaller half.

\paragraph{Why BM25 is weak here.}
Two reasons, both structural. IDF is estimated over the $\sim$24 pages of a
single paper, where term statistics are degenerate; and every page of a paper
shares its topical vocabulary, so the discriminative signal BM25 relies on is
largely absent. Within-document page ranking is close to a worst case for it.

\paragraph{Fix.}
We compared candidate selectors on the same 97 units; \emph{anchors} here are
the four deterministic lookups defined in \S\ref{sec:anchors}:

\begin{center}
\small
\begin{tabular}{lcc}
\toprule
strategy & recall & pages \\
\midrule
BM25 top-4 (original) & 52.6\% & 4.0 \\
dense top-4$^\dagger$ & 40.2\% & 4.0 \\
RRF \cite{cormack2009rrf} top-4 & 53.6\% & 4.0 \\
anchors only & 42.8\% & 3.7 \\
BM25-4 $\cup$ anchors & 65.5\% & 6.6 \\
BM25-8 $\cup$ anchors & 75.8\% & 9.9 \\
\textbf{all pages} & \textbf{100\%} & 23.5 \\
\bottomrule
\end{tabular}
\end{center}

\noindent{\footnotesize $^\dagger$ dense was scored on the first 2{,}000
characters of each page against BM25's full page, so this row is a lower bound
and not a like-for-like comparison.}

A full paper is $\approx$19k tokens at the median (26k at p90) across the 97
units, estimated at four characters per token, comfortably within the context
budget. Since any change to the selector invalidates the
prompt cache and forces the same full re-run regardless of its size, there is no
reason to buy 65\% when 100\% costs the same. We therefore show the whole paper in document order. BM25 unioned with the anchors (\S\ref{sec:anchors}) survives only as a fallback
for papers above a page cap, which on the test split never triggered: it is a
weak ranker for within-document page selection, for the reasons just given, but
weak is not useless when the alternative is showing nothing.

This is a blunt fix. It is not a better ranking function; it is the observation
that the ranking problem was optional at this document length. Any system whose
documents fit in context can do the same.

Visual page retrieval \cite{faysse2025colpali, cho2024m3docrag} is the stronger
method here on paper, and we did not adopt it: at a ceiling of 100\% already
reached by showing every page, its value would be cost control at scale rather
than accuracy, and it carries a substantial dependency. A controlled comparison
of visual against textual page selection on this task remains untried. The main threat to simply
showing everything is position-dependent degradation in long contexts
\cite{liu2024lost}; our questions are predominantly single-page factoid lookups
at $\approx$19k tokens, the regime where that effect is weakest, but we have not
measured it directly.

\section{What structural parsing still buys}
\label{sec:anchors}

Retrieval of any kind cannot serve questions that address their target by
\emph{ordinal} or \emph{number}: ``the first author of the 24th reference'',
``how many parentheses in equation 6''. A bibliography page shares almost no
vocabulary with such a question, so neither lexical nor dense scoring can find
it. We therefore added four deterministic lookups, which we call anchors. Each
takes a literal string in the question, a table number, an equation number, a
reference position, or a distinctive term, and resolves it against the parse
rather than by scoring:

\begin{enumerate}
\itemsep2pt
\item \textbf{Named object}: ``Table 3'' resolves to the page of that table's
  caption, read from the parse's caption index.
\item \textbf{Equation tag}: ``equation 6'' resolves to pages carrying a printed
  ``(6)'' marker.
\item \textbf{Ordinal reference}: ``the 24th reference'' resolves to entry 24 of
  the parsed bibliography, which we split into an addressable array on its
  printed ``[N]'' markers.
\item \textbf{Caption term}: a distinctive term from the question appearing in a
  figure or table caption resolves to that object's page.
\end{enumerate}

\noindent We built these as a page-selection mechanism, to be unioned with BM25,
and on their own they reach 42.8\% gold-page recall with no ranking and no model
call. Showing the whole paper (\S\ref{sec:pageselect}) made that role almost entirely
redundant: anchors enter selection only for papers above the page cap, which was
1 of 58 predicted papers on development and \textbf{0 of 93 on test}. The move
from 52.6\% to 100\% recall is not attributable to the anchors: 95 of the 97
measurement units fit under the cap and saw the whole paper, and on the two
units that did not (one 43-page paper) the BM25 half of the fallback caught the
gold page on its own, with the anchors independently catching it as well.

What survives is not retrieval but parsing, and it operates in stage 2 rather
than in page selection. First, \emph{content}: for a question about the
bibliography we hand the model the resolved entry, or the whole numbered list,
so that ``the 24th reference'' is an array index rather than a counting exercise
performed while reading. Second, \emph{scored identifiers}: the evidence key
includes \texttt{citation\_id} and \texttt{equation\_id}, which we extract
directly from the question when it states them. Our test submission carries 12
and 6 of these respectively; emitting a page without them mismatches gold by
construction, which is what cost us 0.091 evidence $F_1$ on development (0.456 to 0.365 on the
frozen submission) when the evaluator began scoring those fields, before the
fix.

Anchors 1 and 3 are exact: a caption index and a numbered reference list each
map a name to a single page. Anchors 2 and 4 over-trigger, because the string
they match is not unique to the target. For ``equation 6'' in one development
paper the anchor returns pages 2, 7 and 8 of 24, the gold page being 7, since
``(6)'' also appears as a prose back-reference and inside a table. That cost
prompt space rather than accuracy when anchors fed selection, and is moot now
that they rarely do.

\section{A tuning choice that did not transfer}
\label{sec:setsize}

Gold paper sets on the development split are four papers for 27 of 29
multi-paper questions. Our multi-paper route accordingly padded each
LLM-mapped cluster up to four with dense-kNN neighbours of the mapped papers,
on the reasoning that unnamed comparison baselines share the cluster's topic.
We disclosed this as a development-fit choice.

It transferred badly. On the test split the mapping step returns a median of two
papers per component (0:5, 1:20, 2:22, 3:7, 4:4, 6:3 across 61 components), so
padding invented roughly two papers per question. Paper precision fell from
0.840 on development to \textbf{0.453} on test while test recall was 0.773,
dropping paper $F_1$ to 0.517 and the composite score to 0.489. (That 0.840 is
development \emph{precision} under this padded configuration; it is coincidence
that it equals the final system's paper $F_1$ in Table~\ref{tab:results}, whose
precision is 0.935.)

Removing the padding (emitting only the papers the model mapped) raised test
paper $F_1$ to \textbf{0.762} and the composite to 0.568, the largest single
improvement we made. The shipped system predicts a mean of 1.6 papers per
question on the test split: 41 questions receive one paper, 21 two, 8 three or
four, and 1 seven.

The instructive part is that this was not a development/test trade-off at all.
Sweeping the pad target on development gives $F_1$ of 0.840, 0.831, 0.814 and
0.825 for targets of 1, 2, 3 and 4, with precision falling 0.935 to 0.840. The heuristic was mildly harmful on the split it was fitted to. It had been
introduced alongside several other changes and evaluated only in aggregate, so a
net gain from the bundle concealed a component that was costing points
throughout.

We draw two conclusions. First, ablate individually, not in bundles; a disclosed
tuning choice is not the same as a validated one. Second, benchmark splits can
differ in properties as basic as gold-set cardinality, and a system that infers
set size from the development distribution rather than from its own evidence
will over-predict when they do.

\section{Positional bias in multiple-choice answering}
\label{sec:mcbias}

\paragraph{The problem.} Multiple-choice options are presented to the model as a
labelled list, and a model may prefer a label or a position rather than the
option's content. Accuracy alone cannot distinguish such a preference from ordinary error, because
it reports only how often the answer was right. The distribution of labels the
system chooses can.

\paragraph{Why it matters here.} Our test accuracy of 0.840 looked unremarkable,
but the predicted labels were A 8 / B 8 / C 12 / D 22 over 50 questions, against
a plausibly near-uniform gold: the final option was chosen about twice as often
as chance. The distribution did not change when we corrected the upstream paper
predictions (\S\ref{sec:setsize}), which places the cause in answer synthesis
rather than retrieval.

\paragraph{The fix.}

Because the over-chosen label was also the last-presented option, ``prefers the
letter D'' and ``prefers the final position'' are indistinguishable in a single
ordering. We therefore re-ask each question with the option \emph{texts} moved between
label slots (identity, reversal, and a rotation), keeping the label alphabet
fixed so the output format is unchanged, and take a majority vote. A
three-way split retains the original answer, so a lone dissenting view cannot
overwrite it.

We gated deployment on the development split: multiple-choice accuracy rose
0.854 to \textbf{0.902}, from two changed answers, both away from the
over-selected label and both corrections, with all other metrics unchanged. On
test the method changed four answers, all decided 2--1 by the permuted views,
moving accuracy 0.840 to \textbf{0.920} and the distribution to A 8 / B 9 /
C 13 / D 20. All six corrections across both splits ran in the predicted
direction. We note the sample is small and report it as such.

\section{Auditing grounding}
\label{sec:gate}

A grounded-QA pipeline can produce a correct answer for the wrong reason, from the model's pretrained knowledge of a paper rather than from the retrieved
document. The pool is 2024--2025 machine-learning papers, so parametric recall of
these specific results is a live possibility, not a hypothetical.

We therefore added a pre-emit entailment gate. After evidence validation, each
answer component is audited against \emph{only the raw parsed text of the pages
it itself cites}, no pipeline-derived facts or transcripts. The auditor is the
same model as the rest of the pipeline, prompted with only the question, the
answer and the cited page text, and instructed to treat true-but-unstated
claims as unsupported. This
distinction matters: our earlier verification pass re-read the pipeline's own
transcripts and could therefore only detect self-inconsistency, never
ungroundedness.

On development, 25/41 multiple-choice and 14/26 freeform answers are supported
by their own cited source text, 39 of 67 answer components in total, so a
substantial minority assert something their cited evidence does not carry.

We had expected the verdict to predict correctness, and on an earlier
configuration it appeared to. On the final system it does not. Multiple-choice
answers judged \emph{unsupported} are correct slightly more often (15/16) than
supported ones (22/25); freeform shows a weak effect in the expected direction
(8/14 against 5/12) on 26 observations. We report this as a negative result: the gate detects whether an answer is grounded in its own citation, which is what
we built it to do, but on this data that is close to independent of whether the
answer is right. The two properties are worth keeping distinct.

\paragraph{The metric cannot reward abstention.} A grounding-oriented evaluation
might be expected to reward a system for declining to assert what it cannot
support. This one cannot. Because a blank answer and a wrong answer score alike,
withdrawing an unsupported claim can only convert ``unsupported but correct''
into ``blank and wrong'': enforcing the gate costs multiple-choice accuracy
$0.902 \rightarrow 0.537$. This is arithmetic, not tuning, and it applies to any
system on this metric. We therefore run the gate in reporting mode. We raise it
as a task-design observation rather than a system result: an evaluation that
scores grounding separately from answers still gives a system no reason to
prefer silence over a guess.

\section{Results}
\label{sec:results}

Table~\ref{tab:results} reports the final configuration on both splits; test
scores come from the organisers' online evaluator against hidden labels,
reported there under the team name Everest. The
ranking metric averages paper $F_1$, evidence $F_1$ and an answer score, the last
being the mean of the available answer-type metrics using the \emph{macro} cell
figure.

\begin{table}[t]
\centering
\small
\begin{tabular}{lcc}
\toprule
leaderboard metric & dev & test \\
\midrule
paper precision (macro)      & 0.935 & 0.830 \\
paper recall (macro)         & 0.809 & 0.759 \\
paper $F_1$ (macro)          & 0.840 & 0.762 \\
\addlinespace[2pt]
evidence precision (macro)   & 0.482 & 0.479 \\
evidence recall (macro)      & 0.501 & 0.448 \\
evidence $F_1$ (macro)       & 0.468 & 0.441 \\
\addlinespace[2pt]
multiple-choice accuracy     & 0.902 & 0.920 \\
freeform exact match         & 0.500 & n/a \\
table row $F_1$ (macro)      & 0.533 & 0.421 \\
table cell accuracy (macro)  & 0.374 & 0.242 \\
table cell accuracy (micro)  & 0.370 & 0.287 \\
\midrule
composite                    & n/a   & \textbf{0.577} \\
\bottomrule
\end{tabular}
\caption{Every metric the official evaluator reports, on the development split
and the hidden test split. Test figures are from the online evaluator (team
Everest); the
composite is its ranking metric. A development composite is not comparable
because the development split contains freeform questions and the test split does
not.}
\label{tab:results}
\end{table}

On evidence specifically, the page-selection change (\S\ref{sec:pageselect})
moved development evidence $F_1$ from 0.384 to 0.456 under the evaluator current
at the time, with precision rising alongside recall
($0.367 \rightarrow 0.475$), the additional pages replaced wrong locators
with right ones rather than adding noise.  Evidence remains, in absolute terms, the weakest part of the system. Full
gold-page recall does not imply full evidence credit: the coarse key also
requires the right source type and object id, and it requires the page numbering
of our copy of the PDF to match the revision the annotator used. Recall of the
page is necessary, not sufficient.

\section{Error analysis}
\label{sec:errors}

Retrieval errors are now mostly recall, not precision (0.830 against 0.759 on
test; on development the final system misses 38 gold papers while predicting 9
non-gold ones): where we miss, the mapping step failed to associate a named
method with any pool paper, rather than associating it with a wrong one. The residual cases are
descriptive references without a distinctive acronym, where neither the tiered
name matcher nor lexical retrieval has a strong signal.

Evidence errors fall into three groups. First, version mismatch: the annotator's
locator refers to a document revision we cannot obtain, so a correct object
carries the wrong page. Second, source-type disagreement: we once cited the right page of an equation as \texttt{text\_span} where gold
says \texttt{equation\_algorithm}; because the evaluator keys on source type, a
correct page scored zero. This was fixed with explicit source-type routing
guidance in the evidence prompt, and the final system emits
\texttt{equation\_algorithm} with the correct equation identifier for that
case. Third, text-layer gaps: equation bodies are frequently absent from extraction
entirely, so questions about the rendered form of an equation are unanswerable
from text at any context size. One development question of this kind has been
answered correctly, then incorrectly, then correctly again across revisions of
the answering prompts, while the equation body was absent from the extracted
text throughout. Whatever produces the right answer there, it is not reading the
equation; this is the answer-level counterpart of the finding in
\S\ref{sec:gate} that correctness and groundedness are close to independent.

Answer errors concentrate in tables, our weakest component throughout (cell
accuracy 0.242 macro on test against row $F_1$ 0.421). Row keys are usually
recovered; the values inside them are not. In the development cases we inspected, the dominant causes were numeric misreads
from densely formatted tables and cells whose value must be combined across a
multi-level header. Multiple-choice errors are now 4 of 50 after the
option-order correction of \S\ref{sec:mcbias}; we have no gold access for the
test split and cannot characterise the residue further.

\section{Conclusion}

Three components of this system were losing points without any sign in the
end-to-end numbers. The evidence reader looked adequate while its selector found
the right page half the time. A padding heuristic looked like a reasonable prior
on set size while costing 0.245 paper $F_1$. Multiple-choice answering looked
merely imperfect while systematically over-choosing the final option. Each became
visible only when the stage was measured against a ground truth of its own:
selector recall conditional on correct retrieval, a single-parameter ablation, and the distribution of predicted labels.

The practical recommendation is to measure selector recall before investing in
the reader, and, when documents fit in context, not to select at all.

\section*{Limitations}

Sample sizes are small: 55 development questions and 71 test questions. The
multiple-choice debias rests on six corrected answers in total, and while all six
ran in the predicted direction, six is a weak basis for a mechanism claim.

Much of our remaining evidence error is PDF-version mismatch: gold locators were
annotated against document revisions we cannot always fetch. Measured before the
page-selection change, this accounted for roughly 32 of 108 development misses;
we have not re-derived the share under the final configuration, and we do not
know how much of it is recoverable with better acquisition.

Two error sources we did not address: equation content is frequently absent from
the PDF text layer, making some equation-rendering questions unanswerable without
a vision pass we do not route to; and freeform gold is written as full sentences
while our answers are terse, so exact match penalises answers whose content is
correct. Several components carry development-fit choices, enumerated in the
released reproducibility statement.

The system depends on a closed commercial model whose decoding exposes no seed
control, so exact reproduction requires our released response cache rather than
being achievable from source alone. We did not attempt the optional
\texttt{test\_extra} split, which would require on the order of $10^4$ PDFs.

\section{Reproducibility}
\label{sec:repro}

A pipeline built on a closed model with no seed control cannot be reproduced by
re-running it. We therefore ship a harness, \texttt{reproduce.py},\footnote{Code, committed
predictions, run artifacts and the harness:
\url{https://github.com/aadityc91/littraceqa-groundlm2026}} that checks the
paper's central claims at whatever depth the reader's artifacts allow, and reports
\texttt{skip} rather than silent success when something is missing. With only
this repository and the released dataset it verifies the gold file by hash,
recomputes every development metric in Table~\ref{tab:results} from our committed
predictions using the organisers' evaluator rather than our own, and validates
the submission. Given the parsed PDFs it recomputes the measurement of
\S\ref{sec:pageselect}; given the response cache it replays the pipeline and
requires zero cache misses and byte-identical output. All checks pass at
submission.

Three practices made this possible. \emph{Content-addressed calls}: every model
call is keyed by \texttt{sha256(prompt)}, so the cache is an exact record of the
run and a changed prompt invalidates cleanly. \emph{Provenance stamping}: each
run writes an immutable copy of its predictions and metrics tagged with model,
cache generation and git revision, marked dirty when the tree has uncommitted
changes. \emph{Pinning measurements to artifacts}: a measurement's meaning can
depend on upstream state, and after we changed stage 1 the page-selection script
silently returned 90 units and 0.533 rather than 97 and 0.526. The experiment is
now pinned to the stage-1 artifact it was taken from. A reproducibility claim should name the artifact, not only the script.

\bibliography{custom}

\end{document}